# Title: Surveying the side-chain network approach to protein structure and dynamics: The SARS-CoV-2 spike protein as an illustrative case

Anushka Halder[1], Arinnia Anto[2], Varsha Subramanyan[3], Moitrayee Bhattacharyya[1*], Smitha Vishveshwara[3*], Saraswathi Vishveshwara[2*]

[1]Department of Pharmacology, Yale University, New Haven, United States

[2]Molecular Biophysics Unit, Indian Institute of Science, Bangalore, India

[3] Department of Physics, University of Illinois at Urbana-Champaign, Urbana, United States

[*]Corresponding authors: moitrayee.bhattacharyya@yale.edu, smivish@illinois.edu, saraswathi@iisc.ac.in

**Abstract**

Network theory-based approaches provide valuable insights into the variations in global structural connectivity between differing dynamical states of proteins. Our objective is to review network-based analyses to elucidate such variations, especially in the context of subtle conformational changes. We present technical details of the construction and analyses of protein structure networks, encompassing both the non-covalent connectivity and dynamics. We examine the selection of optimal criteria for connectivity based on the physical concept of percolation. We highlight the advantages of using side-chain based network metrics in contrast to backbone measurements. As an illustrative example, we apply the described network approach to investigate the global conformational change between the closed and partially open states of the SARS-CoV-2 spike protein. This conformational change in the spike protein is crucial for coronavirus entry and fusion into human cells. Our analysis reveals global structural reorientations between the two states of the spike protein despite small changes between the two states at the backbone level.   We also observe some differences at strategic locations in the structures, correlating with their functions, asserting the advantages of the side-chain network analysis. Finally we present a view of allostery as a subtle synergistic-global change between the ligand and the receptor, the incorporation of which would enhance the drug design strategies.

## A. Introduction

The concept of allostery has evolved for more than half a century (Monod et al., 1965; Koshland et al., 1966; Cui and Karplus, 2008; Motlagh et al., 2014; Cooper and Dryden, 1984; Changeux, 2011). This word in simple terms means 'action at a distance' and implies long distance communication within and across the three-dimensional structures of proteins. Fundamental understanding of the principles guiding allostery in proteins came from two classical models, the concerted Monod-Wyman-Changeux (MWC) (Monod et al., 1965) model and the sequential Koshland-Nemethy-Filmer (KNF) model (Koshland et al., 1966), with the structural insights coming from one of the earliest crystal structures of haemoglobin (Perutz, 1970). An exponential increase in the availability of protein structures in different functional states has improved our comprehension of the phenomenon of allostery (Greener and Sternberg, 2018; Liu and Nussinov, 2016). Studies over the past decades have associated allostery in proteins with accompanying conformational variations. Such conformational changes range from dramatic alterations at protein backbone-level to subtle re-orchestrations involving protein side-chains in the absence of appreciable backbone variations (Tsai and Nussinov, 2014; Bhattacharyya and Vishveshwara, 2011; Salamanca Viloria et al., 2017; Motlagh et al., 2014). It is this latter mode of conformational fluctuations and long-range signaling that are more challenging to capture.

Current advances in both experimental and theoretical techniques have started shedding light into these subtle conformational variations. In particular, long-range molecular dynamics (MD) simulations, providing equilibrium conformational ensembles, have offered extensive computational characterization of the conformational dynamics in proteins/protein-complexes (Lindorff-Larsen et al., 2016; Mysore et al., 2020; Lindorff-Larsen et al., 2010; Karandur et al., 2020). The goal of these studies ranges from understanding the fundamental biophysical principles to more practical applications for drug design (De Vivo et al., 2016; Borhani and Shaw, 2012). Over the past decades, this topic has been extensively discussed in many excellent articles and reviews (Cui and Karplus, 2008; Bhattacharyya et al., 2016; De Ruvo et al., 2012; Verkhivker et al., 2020; Bagler and Sinha, 2005; Ghosh and Vishveshwara, 2007; Astl et al., 2019; Zhang et al., 2020; Zhang and Nussinov, 2019).

Network theory-based analyses of protein structures (Bagler and Sinha, 2005; Atilgan et al., 2012) and dynamics have provided unprecedented insights into the global structural connectivity of proteins and its complexes in the context of allostery and many other biological processes (Atilgan et al., 2012; Krieger et al., 2020; Verkhivker et al., 2020; Bhattacharyya et al., 2016; Gadiyaram et al., 2019; Di Paola et al., 2013). When combined with information obtained from conformational variations as obtained from molecular dynamics, network approaches have elucidated several examples of protein structure-function relationships (Bhattacharyya et al., 2013; Papaleo, 2015; Doshi et al., 2016; Tse and Verkhivker, 2015; Doruker et al., 2000; Sethi et al., 2009).

In essence, it has become possible to obtain a better perception of biological phenomena at molecular level, such as allostery, evolutionary effects, transport phenomenon, mediated through macromolecules, by employing two major concepts: (1) viewing macromolecules, such as proteins, as one single connected entity, where perturbations can affect the conformations at the local or at the global level, and (2) considering the dynamically accessible conformations of proteins and the inter-conversion of their populations under different conditions as a key to biological functions. Regarding the concept of viewing proteins as a single unit, the connections at the non-covalent level play an important role since these are pliable for minor perturbations that are encountered at normal physiological conditions, unlike the covalently stitched polymer chains.

A number of approaches available in the literature (some of them referenced above) differ in how we view the protein structure as a single unit, connected through non-covalent interactions. One can focus on backbone connectivity alone, or connectivity at the level of side-chains (explicit atoms or a representation through centroids) (Greene, 2012; Kayikci et al., 2018; Bhattacharyya et al., 2016). There are a number of ways to define the connectivity criteria and assign strengths of interactions. Similarly, the dynamical conformational landscape can be obtained at the explicit atomic level or indirectly achieved through methods like ENM, which provides cooperative modes of motion (Krieger et al., 2020; Zhang et al., 2020). The atomic level description can be obtained experimentally through biophysical techniques such as X-ray crystallography, cryo-EM, NMR, and computationally through molecular dynamics (MD) simulations.

The identification of specific regions responsible for the overall perturbation and the re-organization of interactions to yield a different conformation in the landscape has received great attention. This is a crucial step in the process of making the connection between molecular events and protein functions. The methods range from direct analysis of the structures to the ones developed based on the physical, mathematical, and engineering principles. Many such concepts are integrated together in computational programs to obtain critical biological insights

(Verkhivker et al., 2020; Zhang et al., 2020; Greener and Sternberg, 2018). Network theory is a widely used approach which provides explicit information on the role of constituent amino acids on the stability of structure networks at a global level (Gadiyaram et al., 2018; Atilgan et al., 2010; Brown et al., 2017). A vast range of experimental and computational studies have taken up the challenge of correlating biological cellular functions to the molecular level changes. Understanding protein connectivity and dynamics can provide molecular mechanistic insights into the various biological processes, like allostery (Verkhivker et al., 2020; Atilgan et al., 2007; Wang et al., 2020a), protein-protein or protein-nucleic acids interactions (Brinda and Vishveshwara, 2005; Keskin et al., 2005; Sathyapriya et al., 2008; Sethi et al., 2009), and ligand/perturbation-induced conformational variations (Csermely et al., 2013; Bandaru et al., 2017; Creixell et al., 2018). Such calculations may also aid the identification of epitopes for drug-binding and capture drug-induced conformational changes in proteins and protein-complexes (Csermely et al., 2013; Krieger et al., 2020).

The focus of this review is to provide a brief account of the different network theory-based techniques targeted at i) characterizing protein structures as a single entity connected by non-covalent interactions, and ii) integrating with conformational dynamics, for which several comprehensive reviews are available (Verkhivker et al., 2020; Hu et al., 2017; Atilgan et al., 2012). The main emphasis here is on the development and application of protein side-chain network approaches (Bhattacharyya et al., 2016; Salamanca Viloria et al., 2017; Kayikci et al., 2018), which have been shown to capture subtle conformational differences that are sometimes elusive to conventional analyses, such as the root mean square deviation (RMSD) at the backbone level. Here, we have considered the SARS-CoV-2 Spike glycoprotein (Zhu et al., 2020) as an illustrative example to demonstrate the capabilities of side-chain network studies. Our focus on analysing SARS-CoV-2 in particular stems from its critical role in COVID19 and the immediacy posed by the global pandemic caused by this highly infectious coronavirus.

In order to appreciate the relevance of side-chain network studies on the SARS-CoV-2 spike protein, here we provide an introduction to this protein in the context of its structure-dynamics and function. SARS-CoV-2 belongs to the family of β-coronaviruses and is closely related to the earlier pathogens, such as SARS-CoV, and MERS-CoV, which caused severe respiratory diseases in 2004 and 2013, respectively. To develop promising therapeutic strategies, we need a clear understanding of the mechanism of action of the SARS-CoV-2 virus. A succinct summary of the structures of the SARS-CoV-2 spike protein and its interactions with the host cell membrane has been recently provided (Zhu et al., 2020; Xia et al., 2020a; Wang et al.,

2020b). These studies highlight how the recognition of the human ACE2 receptor by the spike protein mediates viral entry into the host cell. A simplified version of the interaction between the human ACE2 receptor and the SARS-CoV-2 spike protein, with an emphasis on the structure of the spike protein, shows the steps that lead to viral fusion to the host cell membrane (Figure 1). Long timescale MD simulations of the viral spike protein in different conformations have been recently made available under the Creative Commons Attribution 4.0 International Public License (D. E. Shaw Research, 2020). Further, a few computational studies on the SARS-CoV-2 spike protein, to explore putative allosteric binding sites (Di Paola et al., 2020) and the role of glycans (Casalino et al., 2020) have also been recently published.

The SARS-CoV-2 spike protein is a homotrimeric complex that is pivotal to the coronavirus entry into host cells and one of the key drug targets for COVID-19 (Hoffmann et al., 2020; Huang et al., 2020). In this article, we have selected the closed (PDB_ID: 6VXX) and the partially open (PDB_ID: 6VYB) structures of the spike protein (Walls et al., 2020) as examples to explicitly elucidate the protein side-chain network concepts. Each subunit in the spike protein is organized into an S1 and S2 domain (Figure 2a-b) (Zhu et al., 2020; Xia et al., 2020a). The S1 domain hosts the receptor-binding domain (RBD) that recognizes the human ACE2 receptor (Figure 2a-b) and the N-terminal domain (NTD). In order to engage the ACE2 receptor, the RBD undergoes a conformational change much like the opening and closing of a hinge (Figure 2c). It is either in the receptor inaccessible state (closed state) or receptor accessible state (open state), governing access to the factors that control ACE2 binding (Figure 2c). The S2 domain hosts the TMPRSS2 cleavage site and the heptad repeat 1 and heptad repeat 2 (HR1/HR2) domains, which are the targets for fusion inhibitors (Xia et al., 2020a, 2020b).

Backbone alignment of a closed structure (PDB_ID: 6VXX) and a partially open structure (PDB_ID: 6VYB) reveals small structural differences except that in the partially opened state the receptor binding domain of one subunit swings outward as compared to the closed state (Figure 2c). This is an ideal model system to apply protein side-chain based network calculations, as the observed backbone changes are small, but carries important functional information. The availability of long-scale MD simulation trajectories of the closed and partially open states of the spike protein (D. E. Shaw Research, 2020) further emboldened our choice of using the spike protein as our model system. This data allows us to demonstrate the capabilities of the dynamically stable protein side-chain network, correlating structural connectivity with conformational dynamics.

In Section B, an overview of network theory in the context of protein side-chain network, connectivity criteria for the protein backbone (PBN) and the side-chain (PScN) networks, selection of optimal strength of interaction from percolation theory perspective, and the cluster identification from graph spectral analysis are presented. In Section C, the method for integrating the network analysis with dynamically stable conformational landscapes is introduced. And the differences between the closed and the open trimeric states of the SARS-CoV-2 spike protein are elucidated through chosen network metrics. A Summary and Outlook is presented in Section D.

## B. Protein Structure Network (PSN) perspective into structural organisation

### *B.1. Network theory-based representation of protein structures*

The overall shape of protein structures at a molecular level is captured elegantly through secondary structures, such as helices, beta strands and sheets, and loops, formed by the backbone atoms of the polypeptide chain. Based on the non-covalent interactions, the Ramachandran map characterized the allowed regions of the backbone torsion angles ($\phi,\psi$) and demonstrated that the allowed conformational space of the polypeptide chain mainly consists of the compact secondary and super-secondary structures, stabilized by backbone hydrogen bonds (Ramachandran et al., 1963). However, there are also numerous examples where despite insignificant differences at the protein backbone level, subtle conformational changes at the protein side-chain level guide a plethora of biological functions (Ghosh and Vishveshwara, 2007; Sethi et al., 2009). Such examples have motivated the development of techniques to completely map structural connectivity of proteins at both the backbone and at the side-chain interactions levels, enabling us to correlate even subtle structural variations that elude backbone-based alignments with biological functions.

The study based on the mathematical principles of network theory enables us to view the protein structures with non-covalent interactions as a single, global network. Numerous studies have formulated the backbone and the side-chain structural connectivity in proteins using adjacency matrices and analysed them using various network metrics (Bhattacharyya et al., 2016; Verkhivker et al., 2020; Kayikci et al., 2018; De Ruvo et al., 2012; del Sol et al., 2006; Astl et al., 2019; Krieger et al., 2020). While protein backbone networks (PBN) capture the non-covalent connectivity at the level of the backbone atoms, protein side-chain networks (PScN/PSN) capture the structural connectivity at the level of non-covalent interactions between side-chain atoms. A representation of the global connectivity across the protein structure in terms of networks can capture the effect of perturbations at the local level and also across the entire protein structure network. This property is key to the understanding of how ligand or mutation-induced local conformational changes affect the global structure of a protein, and therefore its function.

An analysis of the metrics from such a connectivity matrix allows the identification of allosteric communication pathways within a protein structure network by affording insights into the interconnected global architecture of proteins. A variety of network metrics can be used to analyse these protein structure networks (both PBN and PScN). The choice of the network metric being used depends on the question of interest. Briefly, metrics such as hubs and clustering coefficient indicate the degree of a residue and its connectivity to neighbouring residues. The percolation behaviour of a network can be captured in terms of clusters and cliques/communities (Deb et al., 2009; Palla et al., 2005). The molecular details of pathways responsible for allosteric signaling can be examined using the algorithms to identify shortest paths of communication (Tse and Verkhivker, 2015; Ghosh et al., 2011).

### *B.2. Protein Structure Network based on backbone (PBN) and the side-chain (PScN) connectivity*

Protein Structure Network refers to the representation of the three dimensional structural connectivity in a protein in terms of a connectivity or adjacency matrix. In network theory language, individual amino acid residues are termed nodes and the connections between them are defined as edges. In case of the protein backbone network (PBN), Cα atoms are generally considered as the representative of nodes and a distance of about 6.5Å or less (based on the radial distribution of Cα atoms in protein structures) between any two sequentially non-adjacent residues are considered as an edge (Miyazawa and Jernigan, 1985; Patra and Vishveshwara, 2000). The construction and application of PBN has been extensively discussed in earlier reviews (Greene, 2012; Di Paola et al., 2015). Here, our focus is on the technical details of construction and the subsequent application of amino acid side-chain-based protein structure networks denoted as PScN (or PSN). There are different ways of measuring side-chain connectivity, such as the distance between the centroids of the side chains, or all atom-atom pairwise distances. Pairwise distances between all atoms of the side-chain of residues i and j ($n_{ij}$) and values within a distance of 4.5 Å (related to the sum of atomic radii (Heringa and Argos, 1991)) capture explicit atomic level connectivity. A normalization value of the total number of interactions ($n_{ij}$) with respect to the maximum values ($N_i$ and $N_j$) observed from a large dataset of high-resolution crystal structures of proteins provides a uniform basis of evaluation as shown in (Eqn. 1) (Kannan and Vishveshwara, 1999; Sathyapriya et al., 2008).

$$I_{ij} = \frac{n_{ij}}{\sqrt{N_i N_j}} * 100 \qquad \text{.... (Equation 1)}$$

This expression allows us to weigh the strength of the interactions (edge weights) in a systematic manner, which can be uniformly applied to all protein structures. Edge weight ($I_{ij}$) can range from a value of zero to one. Values close to zero and one represent weak and strong side-chain connections, respectively. In general, strong connections can be related to nucleation centres formed by the interactions between the residues pairs such as, oppositely charged, stacked aromatic residues, or polar residues involved in hydrogen bonds. The weak interactions on the other hand, usually emerge from a smaller number of non-covalent interactions ($n_{ij}$) between pairs of hydrophobic amino acid residues. Generally these interactions aid in bridging the strongly connected nucleation centres and in organizing the overall tertiary structure of the proteins. A PScN is constructed based on a user-defined value of $I_{ij}$ (termed $I_{min}$) and an edge is drawn when the calculated $I_{ij}$ between a pair of residues exceeds $I_{min}$.

### *B.3. Percolation profile for the largest connected sub-network as a function of edge weight*

To formalize the effect of the edge weight cut-off ($I_{min}$) on the properties of PScN for protein structures, the concept of the largest connected sub-network (cluster or cliques/communities) transition profile was established and has been applied to a wide range of biological problems (Brinda et al., 2010; Deb et al., 2009). The PScN constructed from low values of $I_{min}$, results in a dense matrix with a large fraction of the residues in the protein getting connected yielding the largest cluster of the size approximately 80-90% of the amino acid residues in the protein. The PScN at higher $I_{min}$ values consists only of strongly connected edges, leading to a sparse matrix. The largest cluster in such a case does not cover a major fraction of the residues in the protein structure. On the other hand, the largest cluster from a PScN of low $I_{min}$, although encompasses a large fraction of residues, the ratio of signal/noise is low in this network. It is therefore important to identify an optimal $I_{min}$ to construct the PScN, without losing information from a sparse graph or encountering low signal/noise from a dense graph.

The identification of the optimal $I_{min}$ to construct the PScN has been addressed from the concepts of percolation within a system, as defined by percolation transition point. In these studies, the PScN is characterized by a macroscopically connected sub-network obtained from

$I_{min}$, around the transition point. The sizes of the largest cluster ($L_{Clu}$) or the largest clique/community ($L_{Cli}$) in the protein structure network are measured as a function of network connectivity at various $I_{min}$ values. Plotting the values of $L_{Clu}$ or $L_{Cli}$ as a function of $I_{min}$ leads to a sigmoidal curve. The transition point of this sigmoidal curve is identified as the percolation transition point at which a giant connected cluster still permeates the protein structure network. Interaction strength ($I_{min}$) around this transition point balances the problems of identifying non-specific, weak interactions at smaller $I_{min}$ values and discontinuous, sparse network connectivity across the structure at high $I_{min}$ values. From earlier studies, it is shown that generally the transition point occurs in the range of $I_{min}$ values 0.2 to 0.4 (Brinda et al., 2010). This transition point is noted to be a common feature in most protein structures.

In this study, we have analyzed the largest cluster percolation profile for the partially open and the closed states of the SARS-CoV-2 spike protein. The plots of $I_{min}$ versus $L_{Clu}$ are generated from the dynamically stable adjacency matrices (the generation of which is described in section C) corresponding to the closed (PDBID: 6VXX) and partially open structure (PDBID: 6VYB) of the spike protein (Figure 3). A noteworthy feature is that the profiles of the closed and the partially open states show some differences in the percolation transition point. These differences are specifically located in the transition regions of $I_{min}$ (between 0.2 to 0.3), with the closed state exhibiting a plateau of the LClu consisting of about 2000 residues, whereas, in the partially-open state, the plateau is around a 1000 residue cluster. Structurally, this is reminiscent of the conformational variations between the two states, with more residues held together as the largest cluster in the closed state, in comparison with the partially-open state in the transition region. Based on this study and the earlier ones, we infer that the results obtained from the interaction strengths around the percolation transition point, $I_{min}$ value of 0.3, are more sensitive and also provide a global view of the structural connectivity in proteins. In the analysis described in subsequent sections, we have used an $I_{min}$ value of 0.3 to generate the networks.

### *B.4. Network parameters of the PScN*

With an increase in availability of data in various fields, the advancement in the research area of large-complex network studies has moved in different directions, such as problem and data driven approaches, development of efficient algorithms, and availability of computational packages (Newman and Girvan, 2004; Newman, 2001; Palla et al., 2005). As we have seen above, the crucial input to obtain a solution to the graph is the connectivity or adjacency matrix in which the nodes and edges are defined based on the chosen application. The connectivity matrices (PBN/PScN) can be analyzed using well-established algorithms and network metrics, which can be used to describe various structural and functional properties of the protein. Some of the frequently used network metrics for analyzing protein/macromolecular structures are hub, clustering coefficient, cluster, cliques and communities, and shortest paths of communication rely on well-established mathematical algorithms (Palla et al., 2005; Newman, 2004, 2001; Dijkstra, 1959). Details of these individual parameters and their physical significance have been extensively discussed in past reviews. A brief description of the parameters and their physical significance follows.

*Hubs* represent highly connected residues in the protein structure network, which essentially refer to the degree of a node. In some general networks, the degree of certain nodes can be very large (Newman, 2001; Tsai et al., 2009). However, the degree of any residue in PScN generally does not exceed 10 due to the steric constraints. The hubs are key in maintaining structural stability and information flow in the protein structure network and are often termed as 'hot spots' in the structure (del Sol et al., 2006; Amitai et al., 2004). *Clusters*, commonly identified using the depth first search (DFS) algorithm (Cormen et al., 2001), are a set of residues that are connected in a way that the number of intra-cluster connections is higher than the number of inter-cluster connections, involving these residues. Clusters evaluated at different $I_{min}$ values are used to predict the strength of connectivity within a network as well as studying interfacial interactions in protein complexes (Brinda and Vishveshwara, 2005).

*Cliques* are defined as completely connected subgraphs in a network such that every residue is connected to every other residue in this sub-graph. An assemblage of cliques that share common edges are termed *communities* (Palla et al., 2005). Cliques and communities are used to identify regions of rigidity and higher order connectivity in protein structures (Ghosh and Vishveshwara, 2008). Like $L_{Clu}$, described in Section *B.3*, the largest identified communities ($L_{Cli}$) can also provide insights into the percolation behaviour of strongly connected components within a protein as a function of $I_{min}$ (Deb et al., 2009). Floyd-Warshall and Dijkstra algorithms for computing the shortest paths of communication has been widely used to determine the critical residues involved in allosteric communication in proteins, and for mapping ligand-induced conformational changes (Bhattacharyya and Vishveshwara, 2011; Pandini et al., 2012; Atilgan et al., 2007). The specific choice of a network metric used for analysing a protein structure network is therefore determined by the structural-biological insight we plan to seek. In Section C, we will demonstrate the application of some of these parameters, by comparing the hubs and cliques/communities between the closed and the partially open states of the SARS-CoV-2 spike protein, in their dynamical equilibrium states.

*B.5. Graph Spectra of PScN*

The graph spectral methods based on analysing eigenvalues and eigenvector components of the connectivity matrices is another approach that has been extensively used to analyze protein structure networks (Hall, 1970). Graph spectral analysis on such a network is performed by studying the eigenspace of the Laplacian matrix associated with it. For a network with *n* nodes, the Laplacian *L* is an *n x n* matrix that satisfies equation 2.

$$X^T L X = \sum_{u \sim v} w_{uv} \big(x(u) - x(v)\big)^2 \ \ldots\ldots. \text{Equation (2)}$$

Where, the summation is over every pair of nodes *(u,v)* connected by an edge with weight $w_{uv}$ for some vector *X* in the space of nodes. It is shown that the Laplacian may be expressed in terms of the degree matrix *D* and the adjacency matrix *A* as equation 3 (Hall, 1970; Chung, 1997).

$$L_{uv} = D_{uv} - A_{uv} \qquad \ldots\ldots. \text{Equation (3)}$$

The eigenvalues and eigenvectors of the Laplacian matrix contain information about the connected components or clusters of the network (Gadiyaram et al., 2016). The eigenvector corresponding to the lowest non-zero eigenvalue of the Laplacian, called the Fiedler vector, contains the clustering information. Sorting the Fiedler vector by value (FVC) (the components range from values -1 to +1) identifies nodes that are part of the same cluster. In this manner, all the clusters in the graph, ranging from the largest cluster with maximum number of residues to isolated edges, can be obtained as an analytical solution to the Laplacian matrix of the graph.

We have considered the example of SARS-CoV-2 spike protein receptor binding domain (RBD) (Figure 2a-b) to demonstrate the capability of the graph spectral analysis from the Laplacian matrix. Specifically, we have extracted the clusters from the sorted Fiedler vector of the receptor binding domain (RBD) of the spike protein (PDB_ID: 6LZG), which is complexed with the target ACE2-Receptor (Wang et al., 2020b). The adjacency matrix is constructed as a binary matrix with $I_{ij} \geq 0.3$, with the elements taking values one and zero, respectively. A plot of sorted Fiedler components (FVC) as a function of the nodes (residue details given in Table S1) is shown in Figure 4a. The slope of the FVC plot is also shown in this Figure, which provides a clearer indication of cluster separation (Sistla et al., 2005). The clusters with the number of residues four and above are plotted on the structure of the RBD (Figure 4b). Thus, the graph spectral analysis is a powerful analytical method to extract the clustering information in protein structure networks.

It should be noted that there are limitations of performing graph spectral calculations, such as on large datasets (e.g., long MD simulation trajectories), as they are computationally expensive. However, the method provides unique information which is difficult to obtain directly from other methods. For instance, information can be extracted not only for clusters within a protein, but also on the interfaces between domains in a single protein or across subunits in multimeric proteins (Brinda et al., 2005; Sistla et al., 2005). Graph spectral studies can also be performed on weighted networks, improving the accuracy. Further, it lends itself for quantitative comparison of networks by providing a score and allows us to identify the regions of the network which are dissimilar. A brief review of these aspects has been presented earlier (Gadiyaram et al., 2019).

## C. Protein Structure Network (PScN) for dynamically accessible conformational ensembles

Biological systems exist in a dynamic equilibrium which alters under different conditions of temperature, ionic strengths, and in complex with endogenous ligands/small molecules/drugs/interacting proteins and so on. A glimpse of the accessible conformational landscape can be obtained by studying a large number of experimentally solved structures in different conditions or through long timescale molecular dynamics (MD) simulations. The network properties that we described above for a single structure of proteins can also be extended to study dynamically average properties of conformational ensembles. Depending on the objective, a judicious choice has to be made as to whether to get the averages from all the structural snapshots along the MD trajectory or from selected structures representing various local minima along the trajectory.

### C.1. Dynamically stable protein structure network (PScN) for conformational ensembles

Analyses of protein conformational ensembles have been facilitated by the development of multiple open-source program packages (Bhattacharyya et al., 2013; Brown et al., 2017; Chakrabarty and Parekh, 2016; Felline et al., 2020; Eargle and Luthey-Schulten, 2012) that analyze multiple structural snapshots in dynamically accessed conformational states. The critical element in many of these open-source software is the ability to implement network theory-based calculations to analyze MD simulation trajectories. Some of these packages

(PSN-Ensemble, webPSN v2.0, and NetworkView) also enable the use of residue pairwise interaction energies to weigh the connectivity matrix.

In this review, we discuss the general concepts of network theory-based analysis of protein conformational ensembles, specifically using PSN-Ensemble as an example software package. The basic principles and capabilities of PSN-Ensemble have been described before (Bhattacharyya et al., 2013). Briefly, PSN-Ensemble provides a consolidated and automated analysis platform, bridging network studies with protein conformational dynamics. Taking the coordinates of structural snapshots from conformational ensembles (MD simulations, NMR studies, or multiple crystal structures) as input, the program computes protein side-chain connectivity matrices (PScN). The individual matrices can be averaged by imposing a user-defined cut-off for dynamic stability (say X%). This 'dynamically-stable' PScN retains any interaction that appears in greater than X% of the structural ensemble, highlighting interactions that persist in a user-defined fraction of the ensemble.

Network parameters and metrics, as described in section B, can be used to analyze the dynamically stable PScN. Using the dynamically stable matrix, PSN-Ensemble can compute structural hotspots (e.g., hubs/cliques) and analyse structural rigidity or flexibility (e.g., cliques/communities) (Ghosh and Vishveshwara, 2008; Bhattacharyya and Vishveshwara, 2011), percolation properties of the network (e.g., clusters and largest cluster transition profile) (Deb et al., 2009; Brinda et al., 2010), molecular determinants of allosteric signaling (e.g., shortest paths of communication) (Ghosh and Vishveshwara, 2007), and ligand/perturbation-induced conformational variations (e.g., hubs/cliques/communities) (Sukhwal et al., 2011; Creixell et al., 2018).

Here we provide an example of the application of network theory to analyze MD simulation trajectories. Using PSN-Ensemble, we analyze the long timescale MD simulation trajectories (10 μs each) of SARS-CoV-2 spike protein in the closed and partially open states (D. E. Shaw Research, 2020). Based on the fact that the interaction strength around the percolation transition point is most sensitive while providing a global view of the structural connectivity (Figure 3), an $I_{min}$ value of 0.3 is chosen to construct the PScN for the two states of the spike protein. Further, the dynamically stable PScN for the MD conformational ensemble is computed at a cut-off of 50%. We compared the hubs and cliques/communities from the dynamically stable PScN for the closed and the partially open states of the spike protein. The results of these analyses on the spike protein are summarized in the subsequent sections as an example of network theory-based comparison of different conformational states.

## *C.2 Analysis of dynamically stable metrics of PScN for the closed and the partially open states of the SARS-CoV-2 spike protein*

In this section we compare the side chain network properties related to rigidity/flexibility (hubs, cliques/communities) from the long timescale MD trajectories on the closed and the partially open states of the SARS-CoV-2 spike protein (10 µs each) (D. E. Shaw Research, 2020). In order to engage the host cell receptor (ACE2), the receptor binding domain (RBD) of the spike protein undergoes conformational changes, much like the opening of a hinge (Walls et al., 2020). The closed state of the spike protein (PDB_id: 6VXX) is receptor inaccessible. A partially open structure, with one of the subunits exhibiting RBD opening (PDB_id: 6VYB), is representative of the receptor accessible states of the protein. Using PSN-Ensemble on the MD simulation trajectories, we analyzed the global conformational changes between these closed and partially open states.

The root mean square deviation resulting from a backbone alignment of the closed (PDB_ID: 6VXX) and the partially open (PDB_ID:6VYB) structures of the spike protein is ~0.5 Å. In addition, each subunit in the trimeric spike protein shows less than ~0.5 Å root mean square deviation when compared to each other, either within or between the two conformational states (closed and partially open). This indicates highly symmetric trimeric organization in the starting structures used for the long MD simulations, in terms of backbone superposition. The two states mainly differ in the conformations of the RBD in one subunit, with the rest of the domains relatively unchanged at the backbone level (Figure 2c) (Walls et al., 2020). Subtle conformational changes that differentiate these two states elude backbone-based structural comparisons, which cannot efficiently capture re-orientations at the protein side-chain level. These factors make the SARS-CoV-2 spike protein conformational states an ideal model system to highlight the benefits of side-chain based network (PScN) analysis.

To compute the dynamically stable PScN, we extracted conformational snapshots every 100 ns from the two MD simulation trajectories of the spike protein in its closed and partially open states (D. E. Shaw Research, 2020). About 100 snapshots from each trajectory are used as an input to the software package PSN-Ensemble to calculate network metrics that persist in atleast 50% of the structural ensemble. A comparison of the dynamically stable hub and cliques/communities between the closed and partially open states of the spike protein shows how the conformational change in the RBD leads to global structural rearrangements, which percolates into the membrane-binding domains of the spike protein. Through this highly relevant example, we reaffirm the advantages of using protein side-chain network-based calculations in capturing the changes in structural connectivity and conformational dynamics in proteins, under different conditions of activity, ligand binding, environmental stimulus, and allosteric communication.

### *C.2.1 Comparison of dynamically stable hubs between the closed and partially open states of spike protein*

As mentioned in Section B.4, the residues that form four or more connections with other residues are defined as hubs and these are considered as dynamically stable, if they appear as hubs in at least 50% of the MD simulation snapshots. These hubs are considered structural hotspots. A change in the number and location of the dynamically stable hub residues in the closed and partially open states of spike protein capture the differences in structural connectivity between these two states. The three subunits in the closed and partially open states of the trimeric SARS-CoV-2 spike protein show a total of 187 common dynamically stable hubs (Table S2). The common hubs between the two states represent the structural connectivity in the PScN that remains unchanged between the two conformational states. The three subunits exhibit mostly symmetrical distribution of these common hubs, both in terms of number of hubs as well as the participating residues (Table S2).

The distinctive structural features of the closed and partially open states are shown by the hubs that are unique to each conformational state. A comparison of these unique hubs reveals striking differences between the closed (64 unique hubs) and the partially open states (74 unique hubs). The number and distribution of these dynamically stable unique hubs show large variations between corresponding subunits of the trimeric spike protein, in going from the closed to the partially open state (Figure 5, Table S2). Strikingly, the number and distribution of the unique hubs among the three subunits within a particular conformational state also show significant differences. This indicates asymmetry between the three subunits of the spike protein, both in the closed and the partially open states. This asymmetry is exhibited at the side-chain network level, despite the highly symmetric nature of the spike protein structure in terms of backbone alignment of every subunit with every other subunit (RMSD less than 0.5 Å).

Depiction of these unique hubs on each subunit of the respective conformational states reveals structural rewiring in the entire SARS-CoV-2 spike protein as the RBD goes from the closed to the open conformation. Here, we discuss the differences observed in chain A (detailed differences for all the three subunits are summarized in Figure 5). In chain A, the partially open state of the spike protein shows an increased number of hubs in the NTD as well as in the region connecting the HR1 and HR2 domains (Figure 5). The increased number of dynamically stable hubs suggests enhanced connectivity in these regions as the spike protein transitions into a partially open state. Our results also suggest that the conformational changes in RBD between the two states induce significant reorganization in the dynamically stable PScN. These global side-chain conformational changes are reflected as differences in the distribution of hubs (Figure 5), especially at sites distant from the RBD, despite minimal backbone reorganisation between the two states.

### C.2.2 Comparison of dynamically stable cliques and communities between the closed and partially open states of spike protein

Cliques represent a subset of residues within a protein structure network where each residue is connected to every other residue (Palla et al., 2005). Cliques represent higher order connectivity in a network, highlighting regions of structural rigidity in the context of protein structures. An assemblage of cliques through shared edges/interactions is defined as communities (details in Section B.4). Communities capture the percolation of structural rigidity through the protein structure network. Together, comparison of cliques/communities reflects subtle conformational changes that alter regions of rigidity/flexibility in protein structural organisation.

We compared the dynamically stable cliques and communities obtained from the SARS-CoV-2 trimeric spike protein in the closed and the partially open states (Figure 6, Table S3a-c). The conformational changes that accompany the transition between the two states of the spike protein are reflected by the cliques/communities that are unique to each state. For clarity, we will only focus on the dynamically stable cliques/communities formed at the trimeric interface between the three subunits (also see supplemental Table S3a-c for a comparison of all cliques/communities between partially open and closed states of the spike protein).

Interfacial cliques/communities are an excellent metric to measure changes in connectivity or interaction between subunits for multimeric proteins. In the closed state of the spike protein, a large number of unique interfacial cliques are seen within the RBD of the three subunits (Figure 6, Table S4). A total of 32 unique interfacial cliques are identified in the closed state, involving residues in the RBD, NTD, SD1, S2, and HR1 domains in the spike protein. This suggests a tightly packed trimeric interface in the closed state with rigid connections between the residues across the three subunits. In contrast, only 21 unique interfacial cliques are observed in the partially open state, with marked alterations in the domains participating in the cliques. Only 8 common interfacial cliques are shared between the two conformational states, indicating significant variations in the trimeric interface packing.

Interestingly, most interfacial cliques formed by the RBD residues are lost in the partially open state. This, as expected, may be due to opening of the RBD in one of the subunits in the spike protein, which leads to a weakening of the interfacial connections involving the RBD residues across the trimer. Interestingly, this conformational change percolates to domains that are distant from the RBD, with cliques altering across the entire spike protein (Figure 6, Table S4). A slight increase in the number of interfacial cliques is noted near the HR1/HR2 domain, especially in the region connection the cores of HR1 and HR2.

In this section we have demonstrated the utility of side-chain network metrics like hubs, clusters, cliques/communities by correlating the function of partial RBD opening to global conformational changes at the side-chain interaction level. We have shown that the local conformational changes at the RBD lead to extensive re-orchestration of the entire spike protein side-chain network.

**D. Summary and Outlook**

The term "allostery" was coined more than half a century ago, to characterize the action of proteins away from the classically identified binding site (Monod et al., 1965; Koshland et al., 1966; Changeux, 2011). The mechanism of action was described through lock-and-key or induced fit models. Our understanding of protein structure-function relationship has increased with advancement in structural biology. Today there is an exponential increase of structural data from experiments such as X-ray crystallography, NMR, and Cryo Electron Microscopy (Nitta et al., 2018; Structural biology shapes up | Science | AAAS). In parallel, computational biology has reached a mature level to explore the conformational space of large protein assemblies through MD simulations (Lindorff-Larsen et al., 2016; Mysore et al., 2020; Wang et al., 2019).

The data from experimental structural biology and the MD simulations have become a rich source of information to investigate macromolecular systems in atomic details. Mining such valuable data for protein conformation and dynamics, in order to unravel biological function at a molecular level and provide meaningful and reliable predictions for experimental biologists, has been a challenge. This has led to multidisciplinary approaches and adaptation of different domain expertise to investigate the importance of specific amino acids towards the stability and functions of proteins from various perspectives. Some of the computational concepts and methods that have made their way to address biological systems are network theory, accessible modes, machine learning, percolation phenomenon, in combination with highly valuable chemical and biological inputs.

Here we have presented a focussed review of the protein structures from a network perspective. Specifically, we have focused on the networks of side-chain connectivity to highlight the unique benefits of this approach. We have described the method of quantifying connectivity and identifying optimal connectivity criteria from the corresponding transition point by employing concepts from percolation theory. We have also discussed the global connectivity of the protein side-chains and clustering of interacting residues from the graph spectral perspective. We have pointed out that the highly similar backbone conformations of proteins can host a repertoire of conformational landscapes, which subtly differ in their side-chain interactions. Thus mild perturbations to proteins can lead to side-chain reorganizations that elude backbone-based structural studies and drive allosteric communication. We have briefly touched upon a variety of approaches to investigate allostery, on which excellent recent reviews are available.

Molecular dynamics simulations can yield an ensemble of protein conformations, which can capture both the backbone and the side-chain level differences in interactions. Analysis of MD simulation trajectories using side-chain network formalism provides a global view of protein structural connectivity from a dynamic perspective. We have reviewed the methodology for such integration of MD simulation with network theory-based analyses.

Due to the global pandemic caused by the highly infectious COVID-19, we have chosen the SARS-CoV-2 spike protein as an example to illustrate the dynamic PScN perspective. We have investigated the molecular dynamics trajectories of the closed and partially open states of the trimeric spike protein that are made available by D.E. Shaw Research (D. E. Shaw Research, 2020). Backbone-based structural comparison between the closed and partially open states reveals minimal structural changes. Highlighting the importance of side-chain network analyses, a dynamic PScN-based comparison reveals key differences between the two conformational states. The present investigation highlights the differences at the side chain interaction level, between the two states such as: (1) the differences in the size of the largest connected clusters ($L_{Clu}$) in the percolation transition region, with the closed state being more stable than the open state, and (2) the differences in the network parameters such as hubs, cliques and communities.

A comparison of the network properties of the partially open and the closed forms of the SARS-CoV-2 spike protein reaffirms that different functional states of proteins can adopt very close backbone topology. While substantial side-chain network parameters like hubs, cliques and communities are also common to both the forms, the unique ones are strategically located in various parts of this multimeric protein. For example, the local conformational changes during the RBD opening leads to extensive re-orchestration of the entire spike protein network, more pronounced in the interfacial region of the trimeric contacts. The different functional states are carefully balanced through the re-organization of side-chain connectivity to mediate interactions with the ACE2 receptor, and ultimately viral fusion to host cell membrane. A detailed study of these interactions between the SARS-CoV-2 spike protein and ACE2 receptor or relevant antibodies/drugs from a side-chain network perspective will be the subject of future investigations.

In addition to offering insights into the structure-function correlation in proteins at the side chain connectivity level, the dynamic network-based studies also provide a new perspective of allostery. The flexibility of the protein involved in interaction with ligand/drug or other proteins are as important as their interacting partners, to have a productive signaling output. Allostery should be viewed as a synergistic-global interaction between the ligand (or the environment) and the receptor. The mechanism of long-distance communication involves specific routes and subtle changes in the communication paths, in order to signal at a distance. Analysis of PScN reveals allosteric communication paths via side-chain interactions even without substantial backbone reorganisation. A stimulus at the ligand binding pocket may be transmitted to the desired destination through subtle reorganization of the side-chain interactions that are allowed in the equilibrium dynamical state. Comparison between the two states of the SARS-CoV-2 spike protein reveals significant changes in the hubs and cliques/communities in regions distant from the RBD. The global reorganisation of the side-chain connectivity between the two states of the spike protein could also influence the communication paths within and across proteins. Thus, one can consider the conformational landscape as being made up of various side-chain network paths.

Finally, in the context of treatment of infections, the antibodies, and the vaccines are produced in response to the global topology of the host protein or receptor. They complement the naturally evolved receptor more globally around the binding sites. The drugs on the other hand, which are designed based mainly on the binding site information may not be highly effective. As we have seen here, the binding site residues are held loosely or tightly by the residue clusters, firmly anchoring some of the interacting residues deep within the pocket. The drug developing strategies would benefit by incorporating the side-chain network connectivity information into their design, thus providing a rationale for incorporating the effects of variations in global structural connectivity in proteins.

**Figure Legends**

**Figure 1**: Sars-Cov-2 employs the spike glycoprotein to enter its host cell. The spike protein is composed of two domains, the S1 domain that hosts the receptor-binding domain (RBD), and an S2 domain. The S2 domain arbitrates the fusion of the viral and host cell membranes. Activation of the spike protein happens by cleavage at two sites (S1/S2 and S2') by the Furin TMPRSS2 protease. The spike protein initially binds to the ACE2 receptor on the host cell through its RBD. On activation, it sheds the S1 domain, enabling the S2 to fuse to the host cell membrane. Figure 1 has been adapted from SARS-CoV-2 cell entry: structural & functional mechanism | Abcam (https://www.abcam.com/content/structural-and-functional-mechanism-of-sars-cov-2-cell-entry).

**Figure 2**: Sequence and structural organization of the SARS-CoV-2 spike protein. (a) Domains of SARS-Cov2 spike protein are depicted along with the two critical sites of cleavage. (b) These domains are shown on the structure of the spike protein (PDB_ID: 6VXX, only chain A is shown for clarity). The backbone is represented as a cartoon and the domains are color-coded based on as in Figure 2a. The first residue at the two cleavage sites are highlighted as spheres and labeled. (c) Backbone alignment (chain B only) of the closed and partially open states of the spike protein reveal conformational changes at the RBD (shown by the arrow), with the rest of the domains showing RMSD less than 0.5Å (a low backbone RMSD of less than 0.5Å is also observed for chain A/C between the closed and partially open states).

**Figure 3**: Percolation profile for the largest connected cluster as a function of edge weight for the closed and partially open structures of the SARS-CoV-2 spike protein. The differences between these conformational states manifest as deviations in the connectivity profile in the region around the percolation point (0.2-0.3).

**Figure 4**: Graph spectral analysis of RBD from SARS-CoV-2 spike protein. (a) Side-chain cluster plot of the RBD obtained from the sorted Feidler Vector Components (FVC-at Iij=0.3), Y-axis: Sorted FVC; X-axis: Nodes corresponding to FVC (node details in Table 1). Nodes

with identical FVC belong to a cluster. The red dotted line represents the slope of the FVC plot. The constant zero values of slope correspond to clusters and spikes in the slope indicate the separation between clusters. (b) Representation of the side-chain clusters identified through FVC on the structure of RBD. The clusters are color-coded green (largest), blue (2nd largest), yellow (6 residue cluster), orange (5 residue cluster), and cyan (4 residue cluster).

**Figure 5**: Depiction of dynamically stable unique hub residues in the three subunits (Chain A/B/C) for the (a) closed and (b) partially open conformational states of the trimeric SARS-Cov2 spike protein. The protein backbone is shown in cartoon representation and each subunit is color coded. The unique hub residues in each subunit for the two conformational states are represented as spheres and these residues are labeled.

**Figure 6**: Depiction of dynamically stable unique cliques at the interface between the three subunits (Chain A/B/C) for the (a) closed and (b) partially open conformational states of the trimeric SARS-Cov2 spike protein. The protein backbone is shown as cartoon and each subunit is color coded. The interfacial cliques are highlighted as spheres. A zoomed in view of each interfacial clique is provided with the participating residues labeled.

## Acknowledgements

M.B. and A.H. thanks NIGMS (R00GM126145) for funding. S.V. and A.A. thanks NASI for fellowships. S.V. and V.S. acknowledges the Institute for Condensed Matter Theory at the University of Illinois at Urbana-Champaign.

Figure 1

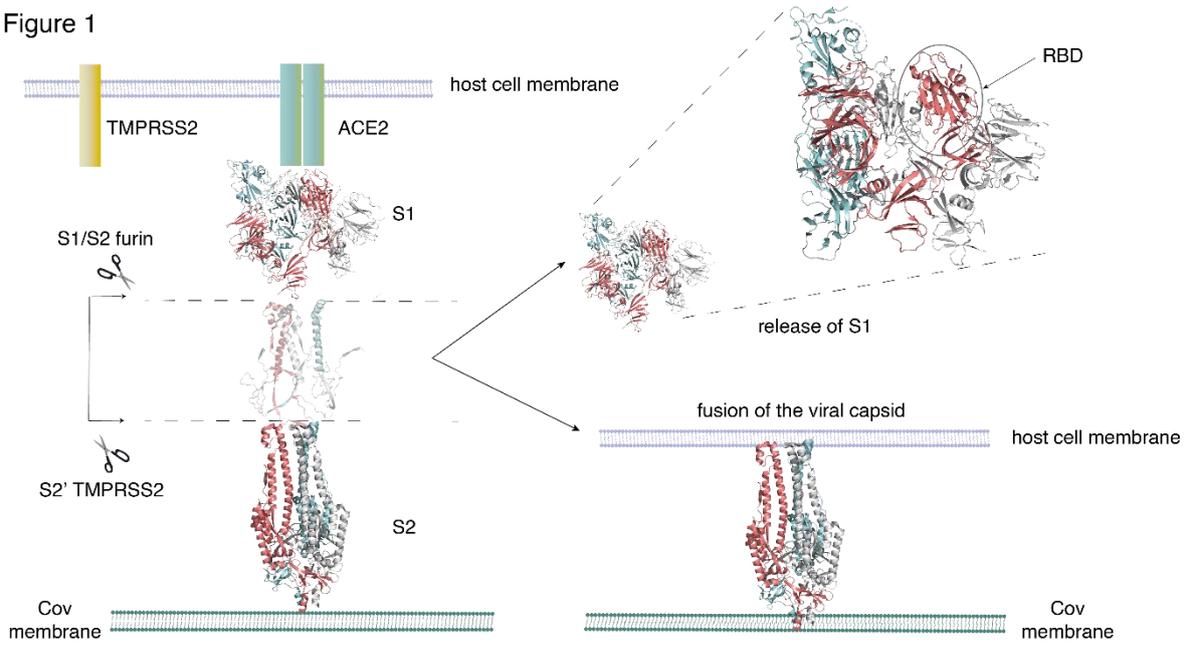

Figure 1: Sars-Cov-2 employs the spike glycoprotein to enter its host cell. The spike protein is composed of two domains, the S1 domain that hosts the receptor-binding domain (RBD), and an S2 domain. The S2 domain arbitrates the fusion of the viral and host cell membranes. Activation of the spike protein happens by cleavage at two sites (S1/S2 and S2') by the Furin TMPRSS2 protease. The spike protein initially binds to the ACE2 receptor on the host cell through its RBD. On activation, it sheds the S1 domain, enabling the S2 to fuse to the host cell membrane. Figure 1 has been adapted from SARS-CoV-2 cell entry: structural & functional mechanism I Abcam. Retrieved July 27, 2020, from https://www.abcam.com/content/structural-and-functional-mechanism-of-sars-cov-2-cell-entry

# Figure 2

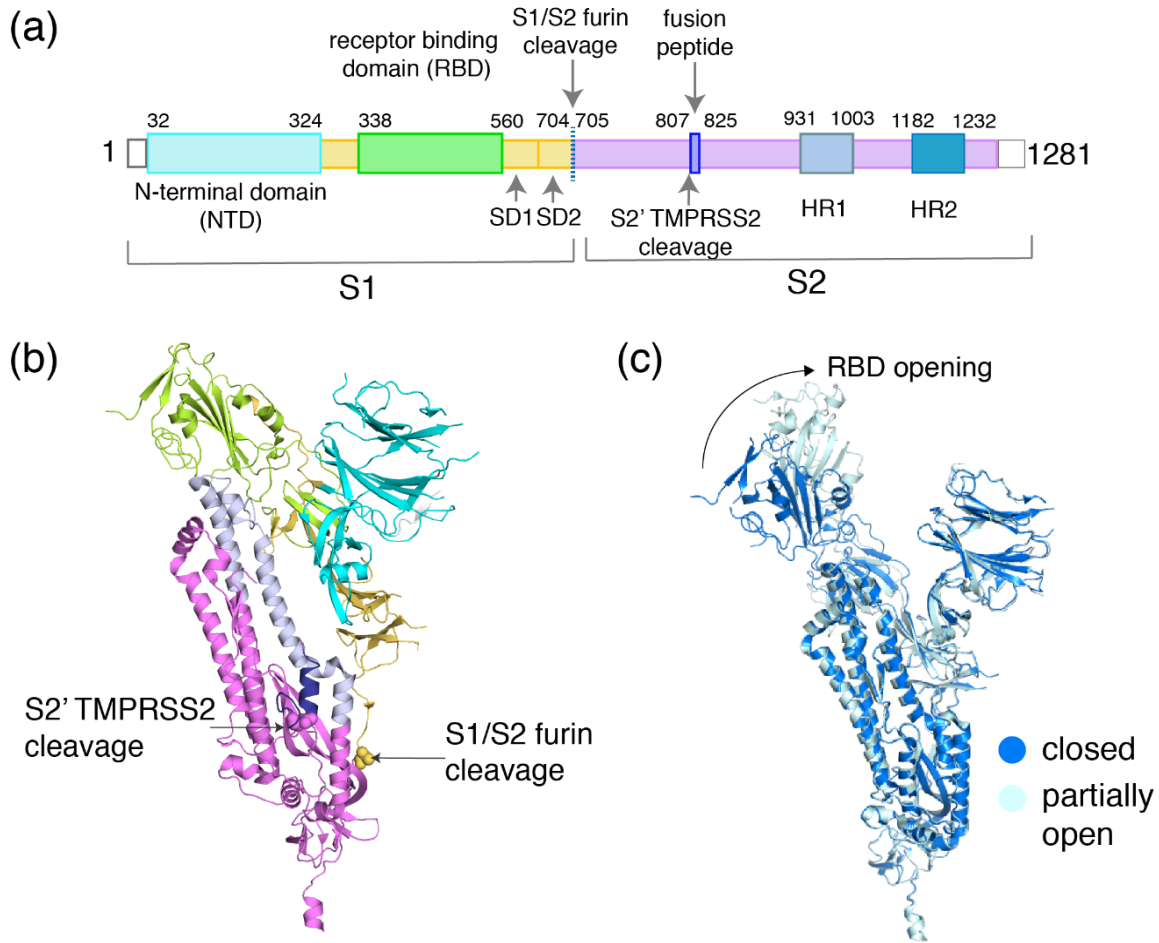

Figure 2: Sequence and structural organization of the SARS-CoV-2 spike protein. (a) Domains of SARS-Cov2 spike protein are depicted along with the two critical sites of cleavage. (b) These domains are shown on the structure of the spike protein (PDB_ID: 6VXX, only chain A is shown for clarity). The backbone is represented as a cartoon and the domains are color-coded based on as in Figure 2a. The first residue at the two cleavage sites are highlighted as spheres and labeled.
(c) Backbone alignment (chain B only) of the closed and partially open states of the spike protein reveal conformational changes at the RBD (shown by the arrow), with the rest of the domains showing RMSD less than 0.5Å (a low backbone RMSD of less than 0.5Å is also observed for chain A/C between the closed and partially open states).

# Figure 3

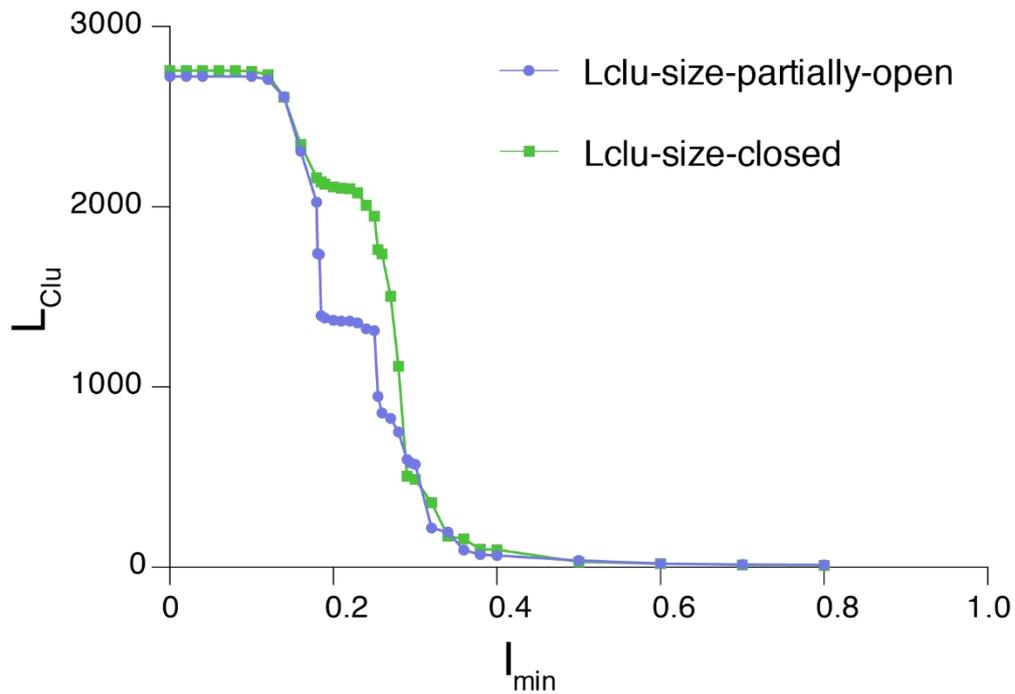

Figure 3: Percolation profile from the largest connected cluster as a function of edge weight for the closed and partially open structures of the SARS-CoV-2 spike protein. The differences between these conformational states manifest as deviations in the connectivity profile in the region around the percolation point (0.2-0.3).

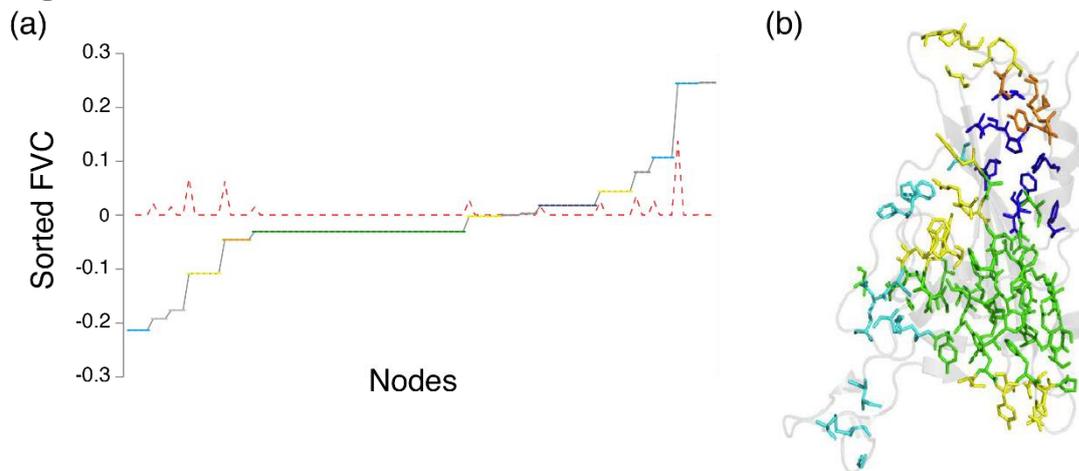

Figure 4: Graph spectral analysis of RBD from SARS-CoV-2 spike protein. (a) Side-chain cluster plot of the RBD obtained from the sorted Feidler Vector Components (FVC-at Iij=0.3), Y-axis: Sorted FVC; X-axis: Nodes corresponding to FVC (node details in Table 1). Nodes with identical FVC belong to a cluster. The red dotted line represents the slope of the FVC plot. The constant zero values of slope correspond to clusters and spikes in the slope indicate the separation between clusters. (b) Representation of the side-chain clusters identified through FVC on the structure of RBD. The clusters are color-coded green (largest), blue (2nd largest), yellow (6 residue cluster), orange (5 residue cluster), and cyan (4 residue cluster).

# Figure 5

(a) Hubs unique to closed state (PDB_id: 6VXX)

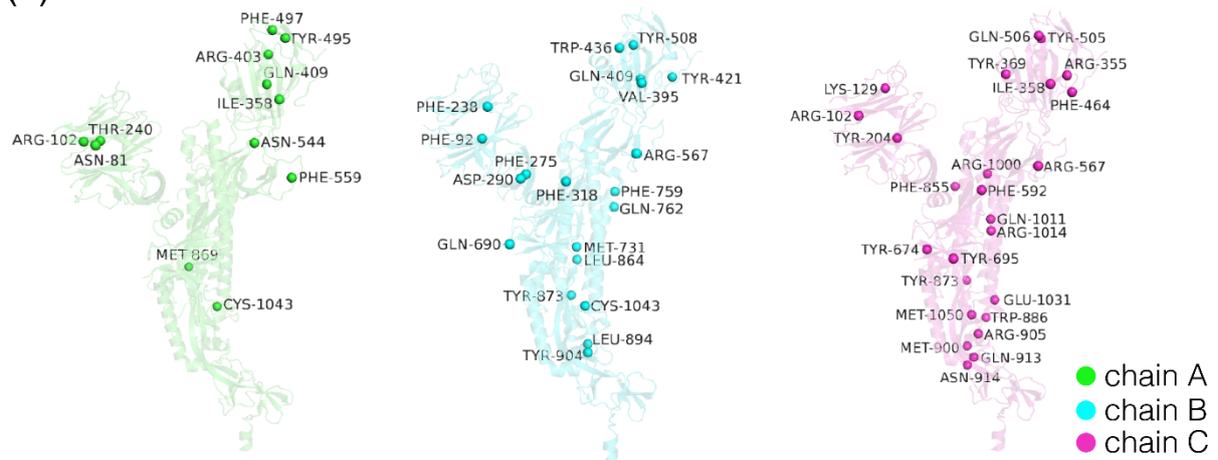

(b) Hubs unique to partially open state (PDB_id: 6VYB)

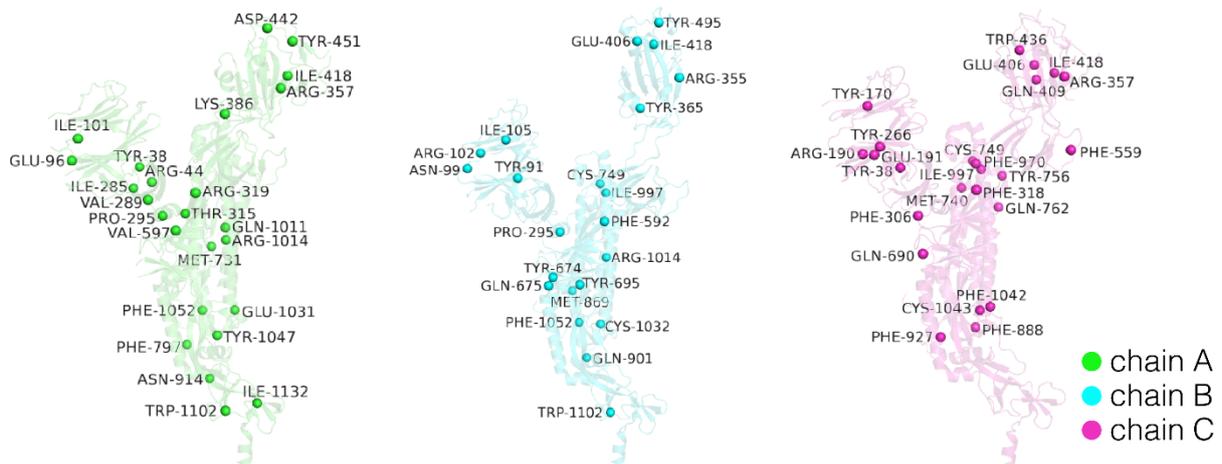

Figure 5: Depiction of dynamically stable unique hub residues in the three subunits (Chain A/B/C) for the (a) closed and (b) partially open conformational states of the trimeric SARS-Cov2 spike protein. The protein backbone is shown in cartoon representation and each subunit is color coded. The unique hub residues in each subunit for the two conformational states are represented as spheres and these residues are labeled.

## Figure 6

### (a) interfacial cliques unique to closed state (6vxx)

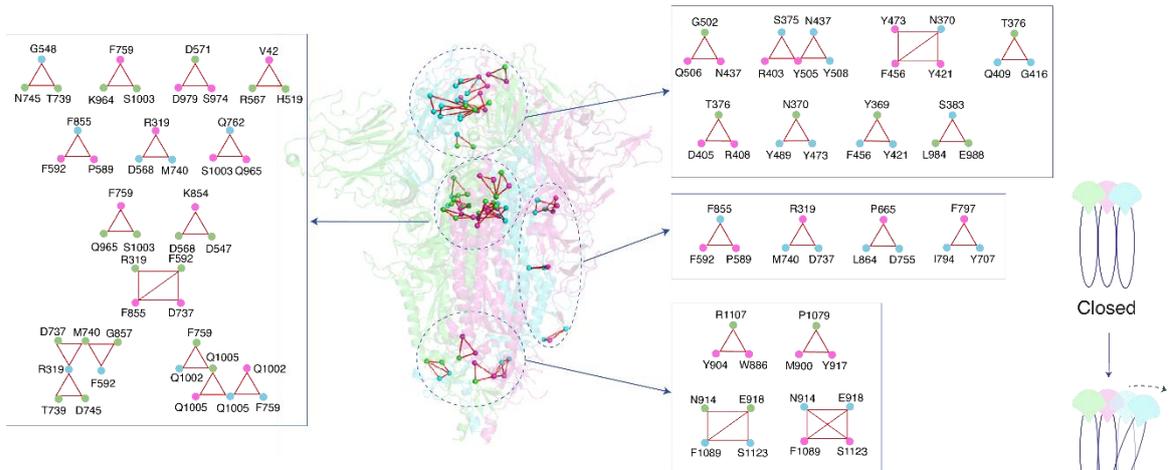

### (b) interfacial cliques unique to partially open state (6vyb)

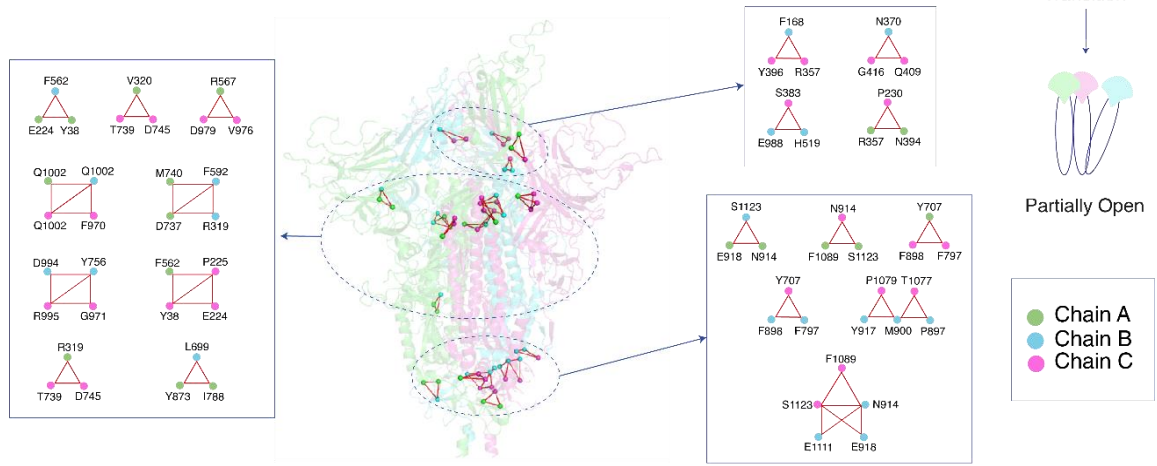

Figure 6: Depiction of dynamically stable unique cliques at the interface between the three subunits (Chain A/B/C) for the (a) closed and (b) partially open conformational states of the trimeric SARS-Cov2 spike protein. The protein backbone is shown as cartoon and each subunit is color coded. The interfacial cliques are highlighted as spheres. A zoomed in view of each interfacial clique is provided with the participating residues labeled.